\documentclass[10pt,aps,pra,twocolumn,superscriptaddress]{revtex4-2}

\usepackage[utf8]{inputenc}
\usepackage[T1]{fontenc}

\usepackage{graphicx}
\usepackage{physics}
\usepackage{amsmath}
\usepackage{amssymb}
\usepackage{xcolor}
\usepackage{comment}
\usepackage{upgreek}
\usepackage{multirow}
\usepackage{hyperref}

\begin{document}

\title{Quantum dynamics for energetic advantage in a charge-based classical full-adder}


\author{João P. Moutinho}
\altaffiliation{Corresponding authors: joao.p.moutinho@tecnico.ulisboa.pt, yasser.omar@tecnico.ulisboa.pt}
\affiliation{Instituto Superior Técnico, Universidade de Lisboa, Portugal}
\affiliation{Instituto de Telecomunicações, Lisboa, Portugal}
\author{Marco Pezzutto}
\affiliation{Portuguese Quantum Institute, Portugal}
\author{Sagar Pratapsi}
\affiliation{Instituto Superior Técnico, Universidade de Lisboa, Portugal}
\affiliation{Instituto de Telecomunicações, Lisboa, Portugal}
\author{Francisco Ferreira da Silva}
\affiliation{QuTech, Delft University of Technology, Lorentzweg 1, 2628 CJ Delft, The Netherlands}
\affiliation{Kavli Institute of Nanoscience, Delft University of Technology, Lorentzweg 1, 2628 CJ Delft, The Netherlands}
\author{Silvano De Franceschi}
\affiliation{Universite Grenoble Alpes, CEA, IRIG, 38000 Grenoble, France}
\author{Sougato Bose}
\affiliation{Department of Physics and Astronomy, University College London, Gower Street, WC1E 6BT London, United Kingdom}
\author{António T. Costa}
\affiliation{International Iberian Nanotechnology Laboratory (INL), Av. Mestre José Veiga, 4715-330 Braga, Portugal}
\author{Yasser Omar}
\altaffiliation{Corresponding authors: joao.p.moutinho@tecnico.ulisboa.pt, yasser.omar@tecnico.ulisboa.pt}
\affiliation{Instituto Superior Técnico, Universidade de Lisboa, Portugal}
\affiliation{Centro de Física e Engenharia de Materiais Avançados (CeFEMA), Physics of Information and Quantum Technologies Group, Portugal}
\affiliation{Portuguese Quantum Institute, Portugal}

\date{\today}

\begin{abstract}
We present a proposal for a one-bit full-adder to process classical information based on the quantum reversible dynamics of a triple quantum dot system. The device works via the repeated execution of a Fredkin gate implemented through the dynamics of a single time-independent Hamiltonian. Our proposal uses realistic parameter values and could be implemented on currently available quantum dot architectures. We compare the estimated ideal energetic requirements for operating our full-adder with those of well-known fully classical devices, and argue that our proposal may provide consistently better energy efficiency. Our work serves as a proof of principle for the development of energy-efficient information technologies operating through coherent quantum dynamics.
\end{abstract}

\maketitle

The ever-growing dependence of society on information technologies has lead to remarkable developments over the past decades. Transistor counts in modern processing devices have roughly doubled every two years, as empirically described by Moore's law \cite{moore1965cramming, kish2002end}, accompanied by similar gains in energy efficiency \cite{koomey2010implications}. However, this exponential increase in computing power is reaching its limits, as miniaturization is becoming increasingly challenging due to thermal constraints \cite{koomey2015primer} and the inevitable influence of quantum effects. At the same time, recent efficiency gains have been mostly due to architecture and algorithmic optimizations, while transistor efficiency has mostly stagnated since the early 2000s \cite{koomey2010implications, koomey2015primer}. This combination of factors is leading computationally intensive tasks to take up an increasing amount of the World's energy budget \cite{masanet2020recalibrating}, accompanied by a similar growth in their carbon footprint \cite{gupta2022chasing}. The energetic cost of information processing has become one of the key challenges to be solved by future generations of information technologies.

Solving these problems may require the development of alternative computing technologies that are both quantum-compatible \cite{benioff1980computer, benioff1982quantum}, and prioritize the usage of energy-efficient building blocks. In this work, we explore this idea by employing reversible quantum dynamics to perform classical information processing. Three-bit logic gates such as the Fredkin and the Toffoli gates are both reversible and universal for classical computation \cite{FredkinToffoli82}, and could act as building blocks for a universal classical machine operating locally through unitary quantum dynamics. Using logically-reversible building blocks opens up the possibility of having a reversible computing machine \cite{Toffoli07, Hanninen13}, thus avoiding the potential source of dissipation associated with bit erasure implied by Landauer's principle \cite{landauer1961irreversibility}.

Our proposal is closely related with quantum computation, a field which has grown significantly in the last decades \cite{preskill2021quantum}. In order to do quantum computation, quantum coherence must be preserved from start to finish, a task which requires the computing device to be well protected from errors, both in terms of isolation from the environment and of the robustness of error-correction protocols \cite{Lidar13}. While many implementations of small-scale devices and algorithms have been demonstrated \cite{Grumbling18, Chuang98}, including some preliminary examples of quantum advantage \cite{Google19, JWPan2020}, these have been limited to less than a hundred noisy qubits and small computation depth. With the current race to achieve full-scale quantum computation, the development of these devices has been fast. Nevertheless, scaling up quantum computing systems is proving to be a challenging problem, and we are still very far from achieving the qubit numbers and robustness needed for fault-tolerant quantum computation \cite{vonNeumann56, Gottesman09}. As a spin-off technology, quantum-based classical computing could prove to be an easier task, differing in two key ways: firstly, it allows binary logic to be encoded in quantum states which do not necessarily form a \textit{quantum bit}, for example different charge states in a quantum dot \cite{divincenzo2000physical}, since there is no need for a coherent superposition of 0 and 1 . Secondly, for an extended computation, quantum coherence is only required at the level of each individual gate \cite{Antonio15}, as the relevant information to be transmitted between gates is entirely classical, and can be limited to computational basis states. Designing energy-efficient quantum-coherent data buses for the transmission of classical data poses also an interesting challenge for quantum-based classical computing, explored in Ref.\ \cite{dylan22}.

As a first step towards the development of a quantum-based classical computer we introduce a 1-bit full-adder composed of a sequence of Fredkin gates implemented in semiconductor quantum dots \cite{patente}, to maximize compatibility with current classical semiconductor technology \cite{gonzalez2021scaling}. Each Fredkin gate is realised by a few electrons evolving coherently under a single time-independent Hamiltonian. Furthermore, we present energetic estimates for our full-adder and discuss how these compare with other proposals.


\section{Background}

\subsection{Semiconductor quantum dots}

Semiconductor quantum dots are one of the many platforms currently being explored for quantum information processing, with various qubit proposals \cite{IEEE85-97,NatNano10-14,li2015conditional,PRApp9-18} relying on the manipulation of charge and spin degrees of freedom of only a few electrons, typically allowing these devices to be small and with fast operation times \cite{NatComm4-13, NatComm5-14}. The large development behind the semiconductor industry that supports our modern day computers is expected to be greatly advantageous to the development of semiconductor-based quantum processors \cite{gonzalez2021scaling}.

Besides the goal of quantum information processing, semiconductor quantum dots have also been used to propose devices for energy-efficient classical information processing based on Quantum Dot Cellular Automatas (QDCA), which have seen continued development over the last two decades. Early experimental results demonstrated the implementation of single QDCA cells and basic logic gates \cite{imre2006majority, amlani1999digital, orlov1997realization}, and since then there have been many proposals for more complex operations, namely full-adders \cite{navi2010new, hashemi2012efficient, taherkhani2017design, abedi2015coplanar}. To the best of our knowledge, there have been no experimental demonstrations of a working QDCA full-adder.

\subsection{Energetics of classical devices}
\label{sec:classical}
In order to benchmark the energy efficiency of our proposal, it is useful to briefly review and estimate the energetic costs of classical information processing devices. Current supercomputer performance and efficiency benchmarks can be found in Ref.\ \cite{Top500}, measured during the execution of large-scale linear algebra tasks. However, our proposal is still restricted to the addition of single bits, and thus it is not directly comparable to \cite{Top500}. Instead, we can use the values therein to get an estimate of the energy cost per bit operation, as detailed in Section I of the Supplementary Information (S.I.). At the time of writing, the estimate for the top performing supercomputer (Frontier) is $1.19\times 10^5$ eV per bit operation, and the estimate for the most energy-efficient supercomputer (Frontier TDS) is $9.95\times 10^4$ eV per bit operation.

Besides the aforementioned estimates for modern supercomputers, there are also numerous proposals for energy efficient full-adder designs based on different technologies. Recent comprehensive reviews of semiconductor-based full-adders were done in Refs.\ \cite{Fulladder2015, hasan2021comprehensive} with detailed simulations of their power consumption, delay, and power-delay product (PDP), which quantifies the total energy spent during one gate operation. Overall, in the analysed proposals, the PDP ranges from $1.8\times 10^2$ to $1.3\times 10^3$ eV for one single-bit addition.

In regard to quantum dot cellular automata, referenced above, simulations of various full-adder designs estimate that their energetic cost could be on the order of the single eV \cite{taherkhani2017design}. However, these designs require dozens of QDCA cells, which is far beyond what has been experimentally demonstrated.


\section{A quantum-coherent classical full-adder}

The full-adder is a basic circuit for addition of binary numbers that can be used to add many-bit numbers when cascaded \cite{Figgatt18}. For this reason, full-adders are a crucial piece of a computer's arithmetic and logic unit (ALU) \cite{Stallings03}. Our proposed full-adder is built via the repeated execution of a Fredkin gate based on the coherent dynamics of a triple quantum dot system. As such we first describe the physical model for the Fredkin gate, and then describe the protocol to operate the full-adder.

\subsection{Fredkin gate model}

\begin{figure}[t]
\centering
\includegraphics[width=0.9\columnwidth]{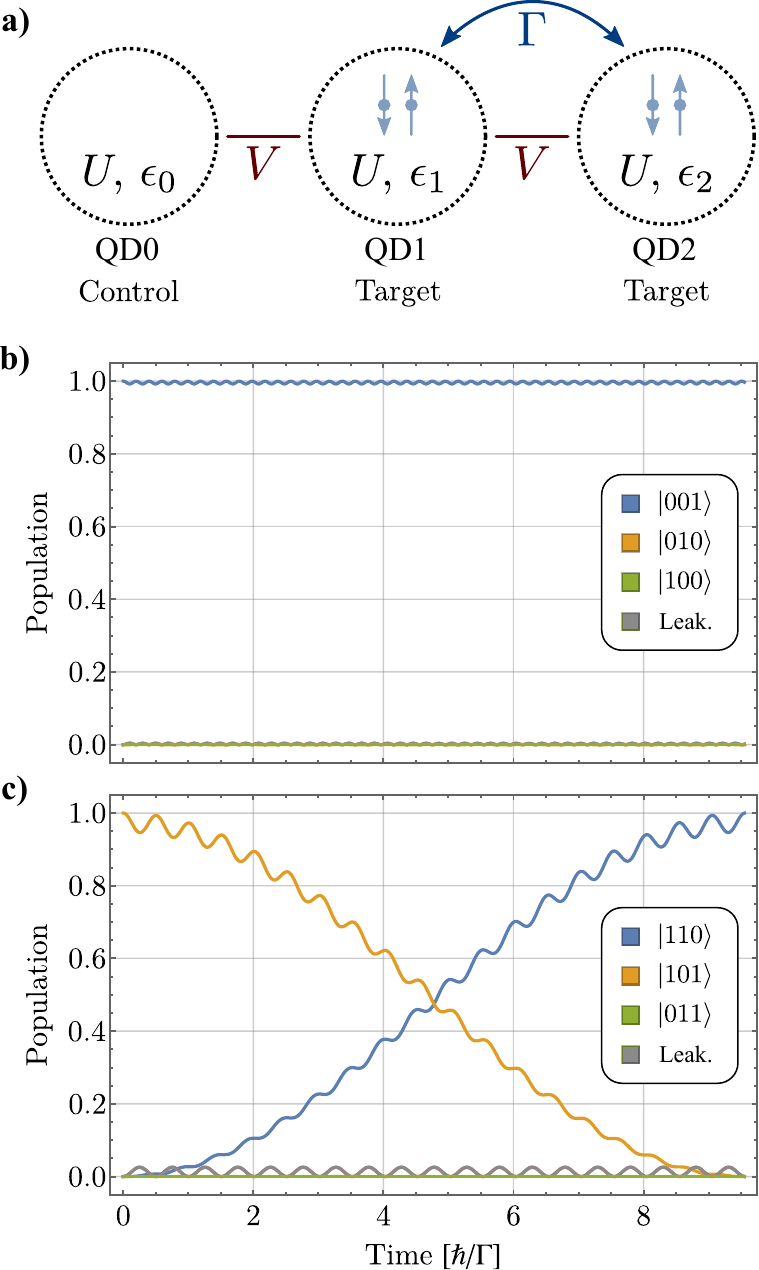}
\caption{\textbf{Fredkin gate dynamics.} \textbf{a)} Interactions represented by the Hamiltonian in Eq.\ \ref{eq:hamiltonian}. By default, we consider the energy levels of QD1 and QD2 to be tuned at resonance ($\varepsilon_1=\varepsilon_2$), allowing the tunneling of electrons. The presence of electrons in QD0 translates to a shift in the energy levels of QD1 produced by the capacitive coupling between these two dots, which detunes QD1 from QD2 thus blocking any coherent tunneling. \textbf{b)}, \textbf{c)} Amplitude squared (population) of logical states as a function of time obtained by the time evolution of the Hamiltonian in Eq.\ \ref{eq:hamiltonian} for two example initial states, $\ket{001}$ and $\ket{101}$ respectively, showing the conditional swap. We label only the relevant logical states, in color, and represent all leakage states in gray. We adopted $\Gamma\sim 44\upmu$eV, $U=21.83\Gamma$, $V=10\Gamma$, $\varepsilon_0=\varepsilon_1=\varepsilon_2=0$, estimates based on realistic values from the experimental literature \cite{simmons2009charge}.}
\label{fig:figure1}
\end{figure}

The Fredkin gate is a three-bit operation, where the logical states of two target bits are swapped if and only if the control bit is in the 1 state. A truth table of this operation is presented in S.I. Table I. We encode each logical state into the charge occupancy of a single-level quantum dot, with an empty dot representing the logical state $\ket{1}$ and a charged dot representing the logical state $\ket{0}$. Note that this encoding does not represent a qubit, as no superposition of $\ket{0}$ and $\ket{1}$ can be prepared \cite{divincenzo2000physical}. Nevertheless, it is sufficient for classical computation. The reverse encoding, with the logical $\ket{0}$ represented by an empty dot, could also be used by appropriately adjusting each dot's on-site potentials. Furthermore, we note that in our model $\ket{0}$ is encoded in a non-degenerate state with either one or two electrons (or, generally, either odd or even electron numbers). Both cases can be represented by the same Hamiltonian and, in principle, the same physical device can realize any of the encodings. Moreover, the choice of encoding is, in practice, implemented by tuning specific parameters in the Hamiltonian (e.g. the strength of a static, uniform magnetic field) and changing details of the control protocol. We will start by presenting our model in terms of the two-electron encoding, and discuss changes resulting from the single-electron encoding. The computational basis is thus defined as
\begin{equation}
\ket{0} = c^\dagger_{\uparrow}c^\dagger_{\downarrow}\ket{\emptyset} \qquad \ket{1} = \ket{\emptyset}
\end{equation}
where $\ket{\emptyset}$ represents the empty QD, and $c^\dagger_{\sigma}$ creates an electron with spin $\sigma$, either $\uparrow$ or $\downarrow$.

We model our device through a general form three-site Hubbard Hamiltonian,
\begin{equation}
\begin{split}
H_F&=\sum_{l\sigma}\varepsilon_l \hat{n}_{l\sigma}
+ \sum_{\sigma}\Gamma(c^\dagger_{1\sigma}c_{2\sigma}+c^\dagger_{2\sigma}c_{1\sigma})\\
&+ \sum_{\sigma\sigma'}V (\hat{n}_{0\sigma}\hat{n}_{1\sigma'}+\hat{n}_{1\sigma}\hat{n}_{2\sigma'} )+\sum_l U_l  \hat{n}_{l\uparrow}\hat{n}_{l\downarrow},
\end{split}
\label{eq:hamiltonian}
\end{equation}
where $l=0,1$ or $2$ is the index of each quantum dot, $\varepsilon_l$ is the single-particle energy of quantum dot $l$, $\Gamma$ is the tunnel coupling between QDs 1 and 2, $U$ is the charging energy of each quantum dot, which we consider equal for simplicity, and $V$ is the capacitive coupling between nearest neighbour quantum dots. We use the operators $c^\dagger_{\sigma}$ ($c_{\sigma}$) for the creation (annihilation) of an electron with spin $\sigma$, and define the number operators as $\hat{n}_{l\sigma}=c^\dagger_{l\sigma}c_{l\sigma}$. We present a schematic of the triple quantum-dot system in Fig.\ \ref{fig:figure1} a). Several experimental works have demonstrated that, by tuning the control gate voltages, it is possible to tune the parameters of quantum-dot arraysd to specific purposes \cite{yu2016tunable, hensgens2017quantum, NanoLett9-09}. We note that this same Hamiltonian has been implemented experimentally in Ref.\ \cite{hensgens2017quantum} using a triple quantum dot system, a concrete demonstration of the feasibility of our proposal. We further discuss experimental details in S.I. Section II E.

A complete analysis of the dynamics implemented by Eq.\ \ref{eq:hamiltonian}, including the effects of quasistatic and high-frequency noise, are presented in S.I. Section II. The Fredkin gate is complete after a time
\begin{equation}
t^*= \frac{2\pi\hbar}{|U-V-\sqrt{16\Gamma^2+(U-V)^2}|}.
\end{equation}
In our simulations we used $\Gamma=44\,\upmu$eV, $V=10\Gamma$ and approximately $U=20\Gamma$, a realistic set of parameters obtained from Ref.\ \cite{NanoLett9-09}. With these values we obtain a gate time of roughly 143 ps. In Fig. \ref{fig:figure1} b) and c) we show a simulation of the time-evolution of Hamiltonian in Eq.\ \ref{eq:hamiltonian} for two initial states, demonstrating the dynamics of the conditional swap between the two target bits. While the time evolution of this Hamiltonian does not implement an exact Fredkin gate, this model allows sufficient flexibility in the parameters to implement a good approximation. For the parameters used, the theoretical fidelity of the gate remains above $0.999$, with quasistatic and high-frequency noise leading to fluctuations in the $10^{-3}$ to $10^{-2}$ range, as described in S.I. Section II A and B.

As mentioned in the beginning, our model also works with a single-electron encoding, in which case we have
\begin{equation}
\ket{0} = c^\dagger_{\sigma}\ket{\emptyset} \qquad \ket{1} = \ket{\emptyset}
\end{equation}
with $\sigma$ fixed to either spin-up or spin-down by imposing a uniform static magnetic field to lift spin degeneracy. While the need for a magnetic field may introduce some challenges when scaling up the system, this encoding may still be more amenable to a near-term experimental implementation given that the swap operation comprises the hopping of a single electron. In contrast to the double-electron encoding, this is faster and less susceptible to dephasing errors. A potential source of dephasing errors is energy relaxation due to electron-phonon coupling. These errors are especially relevant in the two-electron encoding since they will cause two electrons in a single quantum dot to spread to neighbouring quantum dots, leading to leakage states in this configuration. Our estimates for gate speeds, based on typical semiconductor quantum dot parameters, are in the range of hundreds of picoseconds for the double-electron encoding, and tens of picoseconds for the single-electron encoding. Recent experimental studies of coherent quantum dynamics have reported energy relaxation times in the nanoseconds \cite{PRApp9-18}, indicating that the gate implementation we propose here is fast enough to avoid decoherence induced by energy relaxation in both cases.


\subsection{Full-Adder Protocol}

\begin{figure}[t]
\includegraphics[width=1.0\columnwidth]{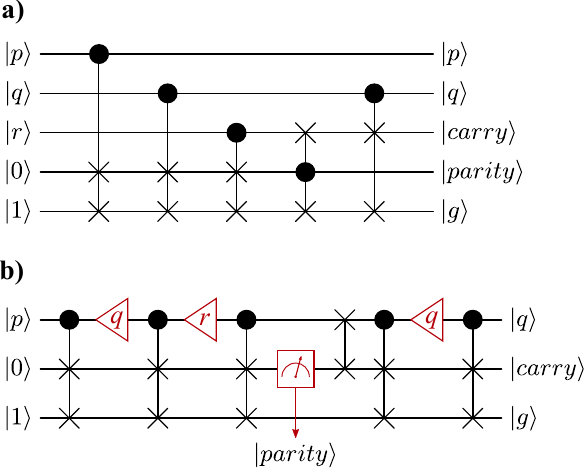}
\caption{\textbf{One-bit full-adder with Fredkin gates.} \textbf{a)} A single-bit full-adder summing bits \textit{p, q} and \textit{r} composed of Fredkin gates can be implemented with two extra ancilla bits initialized in $0$ and $1$. \textbf{b)} To represent the same operation with only three physical bits we collapse the top three lines into a single line representing the control bit for each Fredkin gate. The control bit, first initialized to the logical state $\ket{p}$, must be reinitialized with other logical states during the protocol, marked in the circuit by the left-pointing red triangles. After the third Fredkin gate the second bit must be measured to save the $\ket{parity}$ value, which is part of the full-adder output. Then, a SWAP operation changes $\ket{parity}$ to the top bit to be used as control in the following Fredkin gate. The final Fredkin gate is done after reinitializing the control bit with $\ket{q}$. The output $\ket{carry}$ from the second bit, together with the previous $\ket{parity}$ value, form the output of the full-adder. The extra $\ket{g}$ output serves only to maintain logical reversibility.}
\label{fig:figure3}
\end{figure}

A full-adder sums three bits $p$, $q$ and $r$, outputting two bits: \textit{parity} and \textit{carry}. This operation can be done with a sequence of five Fredkin gates by using two additional bits, as represented schematically in Fig. \ref{fig:figure3} a). A direct, fully coherent implementation of that scheme can in principle be done by coupling at least five distinct quantum dots, each encoding one of the initial $\{p, q, r, 0, 1\}$ logical bits, and then applying the illustrated sequence of five Fredkin gates. Instead, in this section, we focus on a physically simpler setup using only three quantum dots to implement the classical one bit full-adder, following the circuit scheme in Fig.\ \ref{fig:figure3} b). The trade-off is the need to perform at least one intermediate measurement, and the need to reinitialize the control quantum dot with different logical bits. This sacrifice in coherence may incur an additional energetic cost, but the reduced physical system should be advantageous for a near-term experimental demonstration of our work.

A quantum dot implementation of the circuit in Fig. \ref{fig:figure3} b) can be done with minimal changes to the triple quantum dot system used for the Fredkin gate. The key change is the need for electrons to tunnel between QD0 and QD1 in order to implement the intermediate SWAP operation in the circuit. This means that we must modify the previous triple quantum dot Hamiltonian in Eq.\ \ref{eq:hamiltonian} by adding a new tunnel coupling parameterized by $\Gamma^*$,
\begin{equation}
H_A=H_F+\sum_\sigma\Gamma^*(c^\dagger_{0\sigma}c_{1\sigma}+c^\dagger_{1\sigma}c_{0\sigma}).
\label{eq:adder}
\end{equation}
The protocol to run the full-adder with three quantum dots according to the circuit in Fig.\ \ref{fig:figure3} b) is described schematically in Fig.\ \ref{fig:figure4}. It requires a sequence of six steps, after an initial step 0 to load the information for the first Fredkin gate. In the first, second and fifth steps the control quantum dot must be reinitialized with a different logic state, stored in a classical register. Between each step the dynamics are turned on via the tunnel coupling $\Gamma$, allowing the Fredkin gates to operate. During the third step a measurement on QD1 is performed obtaining the logic state corresponding to the parity bit. After the third step the dynamics are activated via the tunnel coupling $\Gamma^*$, instead of $\Gamma$, together with a raise of $2V=20\Gamma$ in the on-site potential $\varepsilon_0$ of QD0, conditional on $\ket{par.}=\ket{1}$. This ensures the execution of an auxiliary swap between QD0 and QD1 independent of the state of QD2. Following this extra operation, the remaining steps are performed as described previously, with a final measurement on QD1 returning the value of the carry bit. Both the parity and carry bits correspond to the output of the full-adder, which can be stored in a classical register after measurement. Scaling up this system may instead shuttle these quantum states directly into the input of a subsequent full-adder. The full scheme and description of the steps can be seen in Fig.\ \ref{fig:figure4}. Fidelity estimates for the proposed full-adder protocol are presented in S.I. Section III showing values close to $0.99$ for all input states, obtained by independent simulations of each intermediate step of the protocol.

\subsection{Energetics}

\begin{figure*}
\includegraphics[width=1.0\textwidth]{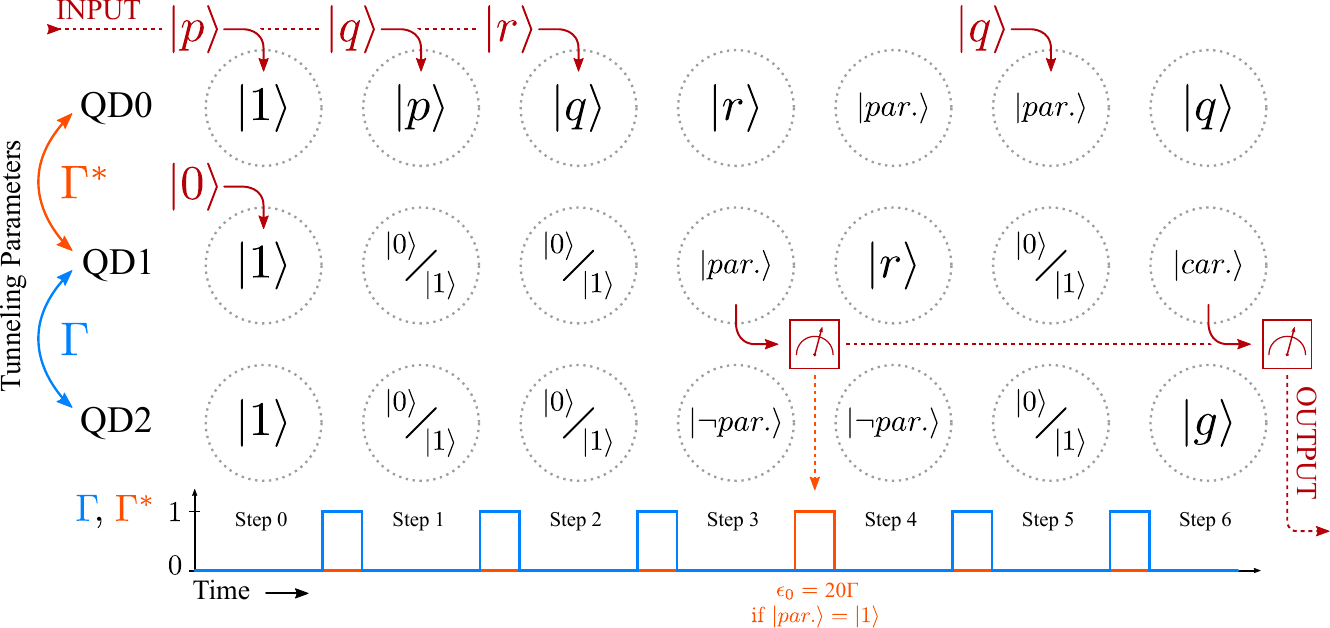}
\caption{\textbf{One-bit full-adder with three quantum dots.} Schematic representation of the logical states in all three quantum dots, where QD0 is represented by the top row, QD1 by the middle row, and QD2 by the bottom row. The seven steps of the protocol are represented from left to right. Considering three initially empty quantum dots, the states $\ket{p}$ and $\ket{0}$ are loaded to QD0 and QD1, respectively (Step 0). The coherent Hamiltonian dynamics are then unfrozen by activating the coupling $\Gamma$ during a time $t^*$ corresponding to the Fredkin gate time. In Steps 1 and 2 the state of QD0 is loaded with $\ket{q}$ and $\ket{r}$, respectively, and two more Fredkin gates are executed. In Step 3, the state of QD1 corresponds to the \textit{parity} bit, which must be measured: if it is 1, the on-site potential $\epsilon_0$ is raised to $2V=20\Gamma$, and if it is 0, $\epsilon_0$ is unchanged. With this condition on $\epsilon_0$, and by then activating $\Gamma^*$, the states of QD0 and QD1 swap independently of the state of QD2. After this auxiliary swap, $\Gamma$ can be activated to complete Step 4. In Step 5 the last Fredkin gate is performed by loading $\ket{q}$ into QD0. Finally, in Step 6, the \textit{carry} bit can be read out of QD1, and the full-adder is complete.\vspace{-7pt}}
\label{fig:figure4}
\end{figure*}

We now wish to estimate the energetic cost of running a full-adder based on our proposed model. The estimates we describe here are based on the theoretical Hamiltonian model used, and in the end we discuss how these may translate to real-world costs. The main cost to consider is the change in the tunnel coupling required to control the flow of electrons between neighbouring quantum dots in each step of the protocol, as illustrated by the pulse sequence at the bottom of Fig. \ref{fig:figure4}. This parameter can be controlled by raising or lowering the potential barrier between two quantum dots which can be described, for example, by a double square-well or a biquadratic potential \cite{yang2011generic}. Given the model described in \cite{yang2011generic}, it can be estimated that a change in the barrier height roughly equal to the charging energy $U$ is enough to change the tunnel coupling from $\Gamma\sim 1$ to $\Gamma\sim 0$ \cite{dylan22}, where effectively no electron hopping occurs between the two dots. As such, for our estimates, we will associate an energetic cost of $U$ to the action of freezing and unfreezing the dynamics. We consider also the worst-case scenario where both increasing and decreasing the barrier will cost the same amount.

Besides the aforementioned costs, we also consider the costs of charging or discharging each quantum dot, having an associated cost $U$ per electron, and the raising of the on-site potentials $\epsilon_l$, costing the amount raised. Analysing the protocol in Fig.\ \ref{fig:figure4} for the operation of a full-adder, we may now count the various actions required at each step and sum their associated costs, as presented in Table \ref{tab:energy}. In this table we include also an estimate for the cost of measurement through a quantum point contact \cite{vandersypen2004real}, $\Delta E_M = 3.6$ keV, as detailed in S.I. Section IV. In contrast to the previous costs, here we used an actual experimental value.

\begin{table}[t]
\centering
\begin{tabular}{c|ccc|c}
\hline\hline
Step & Charging & $~C(\epsilon_l)~$ & $~C(\Gamma)~$ & Measure \\\hline
Step 0 $\rightarrow$ 1 & $4U$ & -    & $2U$ & - \\
Step 1 $\rightarrow$ 2 & $2U$ & -    & $2U$ & - \\
Step 2 $\rightarrow$ 3 & $2U$ & -    & $2U$ & - \\
Step 3 $\rightarrow$ 4 & -    & $20\Gamma$ & $2U$ & $\Delta E_M$ \\
Step 4 $\rightarrow$ 5 & -    & -    & $2U$ & - \\
Step 5 $\rightarrow$ 6 & $2U$ & -    & $2U$ & - \\
Step 6 $\rightarrow$ 0 & $6U$ & -    & -    & $\Delta E_M$ \\\hline
\multirow{2}{*}{Total} & $16U$ & $20\Gamma$ & $12U$ & $2\Delta E_M$ \\
& $15\,\text{meV}$ & $0.9\,\text{meV}$ & $12\,\text{meV}$ & $7.2\,\text{keV}$\\
\hline\hline
\end{tabular}
\caption{\textbf{Full-adder energy cost.} Energy cost of each action required to perform the protocol illustrated in Fig.\ \ref{fig:figure4}. The second column refers to the costs of charging or discharging each quantum dot as $U$ per each potential electron coming in or out of each dot at each step. The third and fourth columns refer to the cost of controlling the parameters $\epsilon_l$ and $\Gamma$ or $\Gamma^*$ of the Hubbard model in Eq.\ \ref{eq:adder}. The final column refers to the cost of measuring the charge state of each quantum dot. The total sum disregarding measurement is $28U + 20\Gamma\sim 28\,\text{meV}$ for $\Gamma = 44\,\upmu$eV, $V=10\Gamma$ and $U=21.83\Gamma$.}
\label{tab:energy}
\end{table}

With the previously used values of $\Gamma = 44\,\upmu$eV, $V=10\Gamma$ and $U=21.83\Gamma$, the estimate for the energetic cost to run the full-adder protocol, disregarding measurements, is $28U+2V\sim 28\,\text{meV}$. This theoretical estimate is significantly lower than the estimate for the cost of measurement, $\Delta E_M = 3.6$ keV, which was obtained from an experimental implementation \cite{vandersypen2004real} where energy efficiency was not a priority. While measurement will certainly incur a substantial cost on a proof-of-principle experiment implementing this protocol, such an action may not be necessary when including such a device in the context of a scalable quantum-based classical computer, where the outputs of single-bit additions could be directly shuttled between registers for more complex operations, for example, by using quantum coherent transport mechanisms \cite{dylan22}. Furthermore, the measurement associated with the auxiliary SWAP after Step 3 is only necessary for our proposed proof-of-principle protocol using three quantum dots, but would not be needed with a more complex setup using five physical bits, directly implementing the scheme in Fig.\ \ref{fig:figure3} a).

The costs listed in Table \ref{tab:energy} can be seen as a theoretical lower bound for the operation of a single full-adder. The costs of actually producing the desired changes in the Hamiltonian of a real system will depend on external factors from the experimental setup used to implement this Hamiltonian, such as the control equipment used, which may have other costs and overheads associated \cite{pauka2021cryogenic}. In the case of gate-defined quantum dots, for example, we can expect that the charging and discharging of the electrostatic gates that produce the desired changes in the Hamiltonian will have energetic costs that may be orders of magnitude greater than the values we have considered.

Regarding external control, some authors \cite{gea2002minimum, gea2003comparison} have argued that, in general, the energy required for this control -- in this case raising and lowering the potential barrier between the quantum dots -- may be significant. This is because the control system needs to have enough energy so that quantum fluctuations of the control and entanglement with the logical system do not have a significant impact on the gate's fidelity.

To the best of our knowledge, this requirement has not been shown to apply universally, and was subject to some debate \cite{itano2003comment, van2003reply, gea2003reply}. In \cite{gea2008quantum}, the authors propose a control scheme that may have a favourable energetic scaling, although it is not clear how to realize such a scheme experimentally. Other strategies may help bring this energetic cost down. For instance, we may find a way to reuse the control energy, by either manipulating several qubits at once or recycling the unused energy.

It is worth noting that similar energetic requirements may also affect conventional computers in the miniaturisation limit, when quantum effects become non-negligible. Therefore, we emphasize that finding an energetically-efficient quantum control scheme is an important open technological problem.

\subsection{Baseline Energetic Costs}

Besides the operational costs to run the proposed protocol for a single full-adder, one should also consider the fixed costs of running a quantum dot processor. We refer to these costs as the \emph{baseline energetic costs}, which include, for example, the cooling equipment, and any room-temperature electronics required. An experimental study of such baseline costs was recently developed for a system of trapped-ions quantum gates \cite{pratapsi22}. These costs can be rather large, on the order of 10 to 15 kW for currently available cryogenic equipment. One of the deciding factors for the success of quantum-based classical computers will be whether or not the baseline energetic costs can scale favourably while the processing power of this new technology is scaled up. One limiting factor could be, for example, the cooling power of the cryogenic equipment. Current fridges typically allow 500 to 1500 $\upmu$W of cooling power at 100 mK \cite{pauka2021cryogenic}. Such a fridge will fit many full-adders running simultaneously as long as the total heat load of the devices and leads connecting to the room-temperature electronics is smaller than the cooling power. If we consider, for example, our ideal 28 meV estimate for the proposed full-adder protocol, a running time of 858 ps corresponding to 6 Fredkin gates, and we assume that all the energy is dissipated as heat, a modern fridge could potentially fit tens of millions of full-adders before surpassing its cooling power. This estimate, while idealistic, indicates that modern cryogenic equipment will be well suited for the continued development of a coherent quantum dot based classical computer, at least in the near future.

\section{Conclusions}

\begin{table}[t]
    \centering
    \begin{tabular}{ccc}
    \hline\hline
    \multicolumn{2}{c}{Technology} & Cost / bit op. (eV)\\\hline
    \multicolumn{2}{c}{Modern Supercomputers} & $\sim 10^5$\\
    \multicolumn{2}{c}{Transistor Full-Adders} & $\sim 10^3$\\
    \multicolumn{2}{c}{QD Cellular Automata} & $\sim 1$\\\hline
    QD Full-Adder & (Coherent) &$\sim 10^{-2}$\\
    QD Full-Adder & (Measurement) &$\sim 10^3$\\\hline\hline
    \end{tabular}
    \caption{\textbf{Energetic cost estimate comparison.} The top three rows are the energy cost estimates per bit operation in modern supercomputers, transistor based full-adders and quantum dot cellular automatas discussed in Section \ref{sec:classical}. In the two bottom rows we present the estimates provided for the operation of our proposed quantum dot full-adder, both for its coherent steps and also including the measurement estimate obtained from experimental literature where energy-efficiency was not a priority.}
    \label{tab:comparison}
\end{table}

In this work we explored the idea of utilizing coherent quantum dynamics to perform energy-efficient classical computation by proposing an experimentally feasible \cite{hensgens2017quantum} full-adder based on semiconductor quantum dots. We present in Table \ref{tab:comparison} a comparison of our energetic estimates with the classical technology introduced in Section \ref{sec:classical}. In summary, our estimates indicate that a coherent charge-based full adder in semiconductor quantum dots may have an ideal energetic cost in the tens of meV range. This would constitute an improvement by several orders of magnitude over the best estimates for the cost per bit operation in semiconductor transistor-based technologies, both in modern supercomputers \cite{Top500} and other full-adder proposals \cite{hasan2021comprehensive}. It should be noted that the estimates used for classical transistor based devices already include implementation specific overheads, while our Hamiltonian based estimates do not. Nevertheless, transistor based devices also benefit from decades of device and architectural developments to optimize both their individual operation and the global control electronics that go into running a fully working processor. In that sense, it can also be argued that any large overheads in the energetic cost of running our proposed device may also be diluted in the future through device and architectural optimizations that will lead to a fully working quantum-based classical processor. At the same time, the development of quantum-compatible classical processing devices may have other advantages such as problem-specific applications in the implementation of hybrid quantum-classical algorithms \cite{mcclean2016theory}, or the development of cryogenic control electronics for quantum-computing applications \cite{pauka2021cryogenic}.

In comparison with quantum dot cellular automata (QDCA), our estimated energetic cost is two to three orders of magnitude better than other estimates in this field. Furthermore, our proposal can be implemented on a triple quantum dot system, which is simpler and has been experimentally demonstrated \cite{hensgens2017quantum}. In contrast, each QDCA cell typically has four quantum dots, and full-adder proposals use 30 to 50 cells each, which is substantially more complex to implement experimentally \cite{navi2010new, hashemi2012efficient, taherkhani2017design, abedi2015coplanar}.

Our approach was to encode the logical state of each bit with one or two electrons in a single quantum dot. While our encoding is not compatible with quantum computing, similar devices exploiting quantum dynamics with the goal of energy-efficiency may also be designed using compatible qubit encodings. Nevertheless, the possibility of encoding logic in more general quantum states eases some restrictions on device design which may be beneficial for the future scalability of such technologies. Besides quantum dots, other platforms currently being explored for quantum computing such as trapped ions or superconductors may also be considered. In fact, a classical half-adder and Toffoli gate using trapped ions was recently proposed and experimentally demonstrated as an alternative for energy-efficient computing \cite{pratapsi22}. In this regard, semiconductor quantum dots may prove advantageous due to their potential compatibility with the already advanced semiconductor fabrication techniques \cite{gonzalez2021scaling}.

Our results are a first step in a roadmap to develop energy-efficient classical computers -- and information technologies in general -- exploiting quantum dynamics and quantum technologies. Several significant challenges will need to be addressed, such as the study of the energetics of quantum technologies \cite{buffoni2022, auffeves2022quantum, montangero2022, pratapsi22}, the development of energy-efficient control systems, compatible data buses \cite{dylan22} and memory devices, and ultimately the combination of all elements into a scalable and energy-efficient processor architecture. Our work proposes a first step in this direction, which may prove to be a promising alternative for the future of high-performance computers when current transistor-based semiconductor technologies reach their absolute limits.

\section*{Acknowledgements}

The authors thank Lieven Vandersypen, Fabian Hartmann, John Quilter, Dylan Lewis, Lorenzo Buffoni and Duarte Magano for helpful discussions and feedback. Furthermore, JM, MP, SP, YO thank the support from FCT -- Fundação para a Ciência e a Tecnologia (Portugal), namely through projects UIDB/50008/2020 and UIDB/04540/2020, as well as project TheBlinQC supported by the EU H2020 QuantERA ERA-NET Cofund in Quantum Technologies and by FCT (QuantERA/0001/2017), and from the EU H2020 Quantum Flagship project QMiCS (820505). JPM acknowledges the support of FCT through scholarship SFRH/BD/144151/2019. SP thanks the support from the ”la Caixa” foundation through scholarship No.\ LCF/BQ/DR20/11790030. SB thanks EPSRC grant EP/R029075/1.





\bibliography{draft}

\setcounter{figure}{0}  
\setcounter{section}{0}
\setcounter{table}{0}
\clearpage

\onecolumngrid
\begin{center}
\large
\textbf{Supplementary Information for}\\
\textbf{Quantum dynamics for energetic advantage in a charge-based classical full-adder}
\end{center}
\vspace{10pt}
\twocolumngrid

\section{Energetics of classical information processing devices}

The energy efficiency of classical information processing devices is measured in performance per watt, which gives the rate of computation that a given device can deliver for every watt of power consumed. The most widespread measures of performance are FLOPS (floating point operations per second) and MIPS (million instructions per second). The power measure can also be defined in more than one way: it can be just the power consumed by the hardware itself, or it may also include the power needed to run any cooling, control or monitoring systems. 

The state of the art of classical supercomputing devices is compiled by the Top 500 project, which ranks supercomputers by performance and computing efficiency \cite{Top500}. The measure of efficiency used is the ability to solve a set of linear equations $Ax = b$, using a dense random matrix $A$. As of June 2022, the most energy efficient supercomputer is the Frontier TDS, with a power efficiency of $62.684$ GFLOPS/Watt. In terms of raw performance, the fastest supercomputer in the world as of June 2022 is the Frontier, with a peak performance of $1,685.65$ PFLOPS and a power efficiency of $52.227$ GFLOPS/Watt.

We would like to benchmark the energy efficiency of our proposal for quantum-based classical computing against the classical state of the art. However, we are still restricted to adding bits, making a direct comparison to the Top 500 project in solving linear algebra problems impossible. As an alternative, we can use the values in the Top500 project to estimate the energy per bit operation of each system. First, we note that the values of GFLOPS/W can be directly used as the number of floating point operations per Joule of energy consumed. We now require an estimate for the number of single-bit operation per FLOP, which is highly dependent on the architecture of each processor. Referring to the discussion in Ref.\ \cite{bitperflop}, it has been estimated based on specific multi-bit adder implementations that a FLOP can take as many as 1000 single bit operations. Using that estimate, a GFLOP equals $10^{12}$ single bit operations. We can now write the total bit operations per Joule and invert to obtain the energy cost per bit operation,
\begin{equation}
\begin{split}
\text{Frontier:}\quad &52.227\times 10^{12}~\text{bit-op/J}\\
\Rightarrow~&1.19\times10^5~\text{eV/bit-op}\\
~\\
\text{Frontier TDS:}\quad &62.684\times 10^{12}~\text{bit-op/J}\\
\Rightarrow~&9.95\times10^4~\text{eV/bit-op}
\end{split}
\end{equation}


\section{Fredkin gate model}

\label{sec:fredkingate}
The Fredkin gate is composed of three bits (two target, one control), that we associate to three single-level quantum dots. The logical states are encoded into the charge states 
of the quantum dots. Although there are different possible choices for this encoding, the Hamiltonian that will implement the gate is the same for all of them. This means, for
instance, that the same physical device can realize any of the encodings. The choice of encoding is, in practice, implemented by tuning specific parameters in the Hamiltonian
(e.g. the strength of a static, uniform magnetic field) and changing details of the control protocol.

We will start by choosing two of the possible encondings. One choice corresponds to associating double occupancy with the logical state $\ket{0}$ and zero charge the logical state $\ket{1}$. The other encoding corresponds to associating the presence of a single electron in a QD with the logical state $\ket{0}$ and an empty QD with the logical state $\ket{1}$. We will first examine in details the consequences of choosing the double-occupancy encoding. In the end we will show the changes resulting from adopting the single-electron encoding. As we will see, the gate operation is qualitatively very similar in both cases, with the main difference being the gate speed, which can be considerably higher for the single-electron case.
In what follows we will refer to the quantum dot associated with the control bit by using the label $0$. The quantum dots associated with the target bits will be labeled $1$ and $2$. The computational basis for each bit is, thus,
\begin{equation}
\ket{0} = c^\dagger_{\uparrow}c^\dagger_{\downarrow}\ket{\emptyset} \qquad \ket{1} = \ket{\emptyset}
\end{equation}
where $\ket{\emptyset}$ represents the empty QD, and $c^\dagger_{\sigma}$ creates an electron with spin $\sigma$. There are two states that are not in the computational basis, namely $c^\dagger_{\uparrow}\ket{\emptyset}$ and $c^\dagger_{\downarrow}\ket{\emptyset}$. This could, in principle, lead to leakage errors, but it is possible to design the dynamics such that this kind of error is negligible, as we shall see in the following.

To obtain an ideal version of the Fredkin gate we look for a Hamiltonian that implements the following time evolutions:
\begin{align}
U(\tau^*)\ket{0}_{0}\ket{a}_{1}\ket{b}_{2} = \ket{0}_{0}\ket{a}_{1}\ket{b}_{2},\\
U(\tau^*)\ket{1}_{0}\ket{a}_{1}\ket{b}_{2} = \ket{1}_{0}\ket{b}_{1}\ket{a}_{2},
\end{align}
where $a,b\in \{0,1\}$. From the time at which the initial state of the gate is prepared ($\tau=0$) to the time of the completion of the gate ($\tau^*$) we would like to use a time-independent hamiltonian. At $\tau^*$ an additional field has to be applied to the system to ``freeze'' it in the desired state.

\begin{table}[t]
    \centering
    \begin{tabular}{ccc|ccc}
    \hline\hline
    \multicolumn{3}{c}{Input} & \multicolumn{3}{c}{Output} \\\hline
    $~C~$ & $~T_1~$ & $~T_2~$ & $~C~$ & $~T_1~$ & $~T_2~$\\
    0 & 0 & 0 & 0 & 0 & 0\\
    0 & 0 & 1 & 0 & 0 & 1\\
    0 & 1 & 0 & 0 & 1 & 0\\
    0 & 1 & 1 & 0 & 1 & 1\\
    1 & 0 & 0 & 1 & 0 & 0\\
    1 & 0 & 1 & 1 & 1 & 0\\
    1 & 1 & 0 & 1 & 0 & 1\\
    1 & 1 & 1 & 1 & 1 & 1\\\hline\hline
    \end{tabular}
    \caption{\textbf{Fredkin gate truth table.} Conditional swap of the target bits ($T_1$ and $T_2$) based on the control ($C$) being 1.}
    \label{tab:fredkin}
\end{table}

\begin{figure}[t]
\centering
\includegraphics[width=1.0\columnwidth]{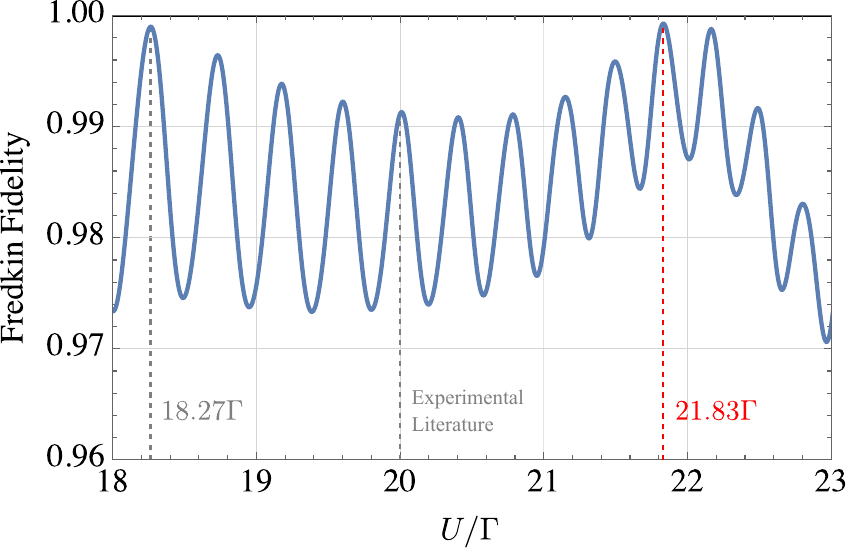}
\caption{\textbf{Change in the Fredkin gate fidelity with the charging energy $U$.} The Fredkin gate fidelity is defined as the sum over the population $|c|^2$ of the output basis states given the time-evolution of each input state, according to Table \ref{tab:fredkin} normalized by $1/8$. Given the literature that identify $U=20\Gamma$ as a realistic value in an experimental quantum dot setting, we choose $U=21.83\Gamma$ as the default value for our simulations to maximize the fidelity. Other parameters are fixed at $V=10\Gamma$ and $\epsilon_l=0$ for $l=0,1,2$.}
\label{fig:fidelity}
\end{figure}

\begin{figure*}
\centering
\includegraphics[width=0.9\textwidth]{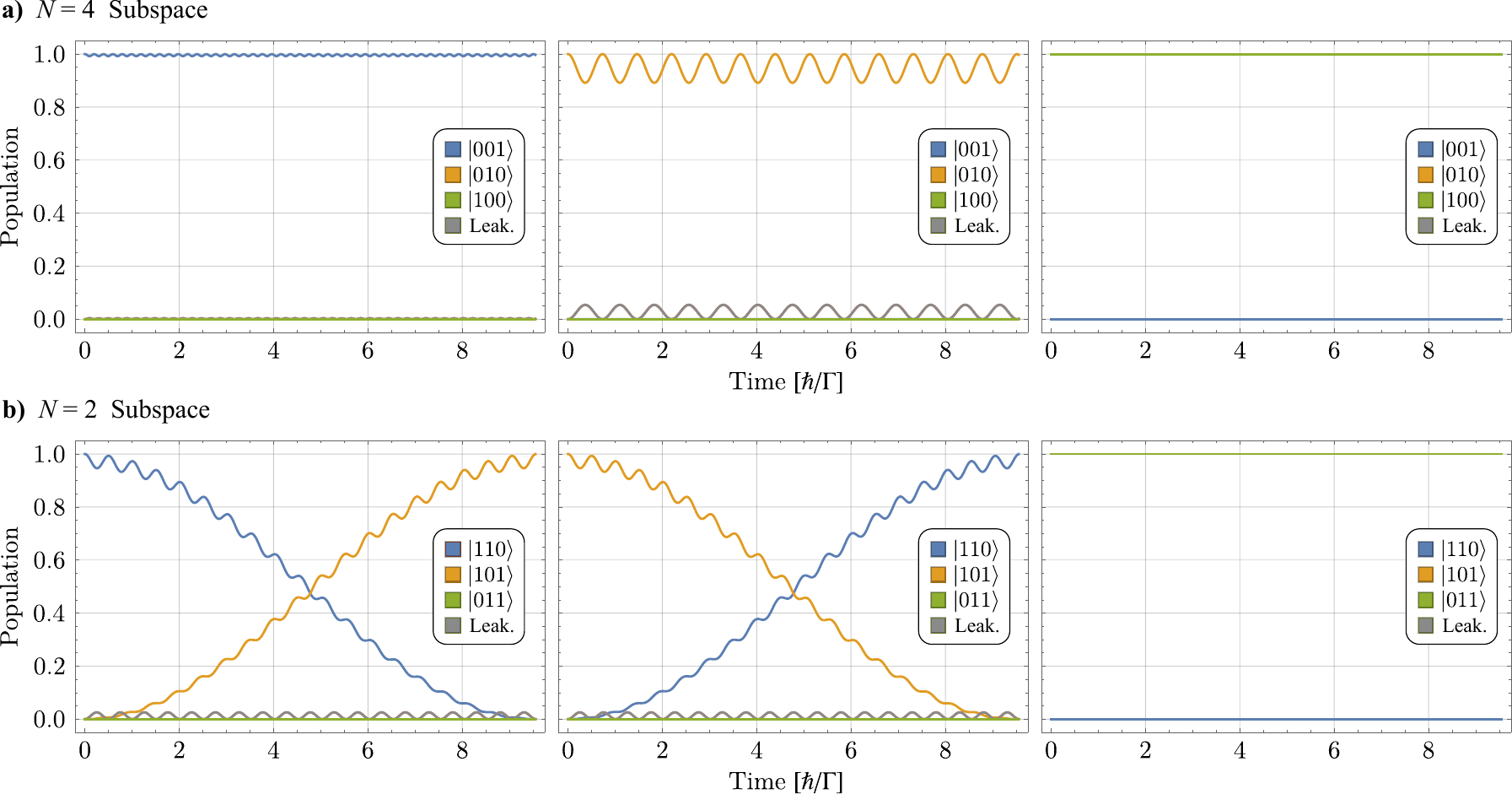}
\caption{\textbf{Time evolution under the Fredkin gate Hamiltonian.} Time-evolution for all logical initial states in the \textbf{a)} $N=4$ and \textbf{b)} $N=2$ subspaces. The $N=0$ and $N=6$ subspaces, not shown, have no dynamics as the quantum dots are all empty or all full.}
\label{fig:swap}
\end{figure*}

In order to design the hamiltonian we first notice that the control bit has no dynamics, irrespective of the gate's initial state. This suggests that the associated QD should be coupled only capacitively to the target QDs. When the control bit is set to 1 the states of QD 1 and QD 2 have to be swapped. Thus there must be a finite tunnel coupling between QDs 1 and 2. From our choice of logical state encoding we immediately see that the total occupancy $N$ of the three dots must be equal to 0, 2, 4 or 6 electrons. For $N=0$ or $N=6$ there are no dynamics, as all dots are either empty or full. This is in accordance with the Fredkin gate rules for the states $\ket{000}$ and $\ket{111}$. To investigate the dynamics for $N=2$ and 4 we have to specify the hamiltonian. We choose the general form
\begin{equation}
\begin{split}
H_F&=\sum_{l\sigma}\varepsilon_l \hat{n}_{l\sigma}
+ \sum_{\sigma}\Gamma(c^\dagger_{1\sigma}c_{2\sigma}+c^\dagger_{2\sigma}c_{1\sigma})\\
&+ \sum_{\sigma\sigma'}V (\hat{n}_{0\sigma}\hat{n}_{1\sigma'}+\hat{n}_{1\sigma}\hat{n}_{2\sigma'} )+\sum_l U_l  \hat{n}_{l\uparrow}\hat{n}_{l\downarrow},
\end{split}
\label{eq:hamiltonian}
\end{equation}
where $\varepsilon_l$ is the single-particle energy of QD $l$, $\Gamma$ is the tunnel coupling between QDs 1 and 2, $U_l$ is the charging energy of QD $l$ and $V$ is the capacitive coupling between nearest neighbour QDs. We also define the number operators $\hat{n}_{l\sigma}=c^\dagger_{l\sigma}c_{l\sigma}$. When the gate is initialized, only states with total $S^z_T=S^z_1+S^z_2=0$ are populated. Since the hamiltonian~\ref{eq:hamiltonian} commutes with $S^z_T$ we only need to investigate the dynamics within the $S^z_T=0$ subspace. Moreover, the two subspaces $\mathcal{H}_c$, $c=0,~1$, characterized by the control QD being in the state $\ket{c}$, are not coupled by $H$, and their dynamics can be analyzed independently.

We choose the 1-2 hopping $\Gamma$ as the energy unit. We also set $\hbar=1$ for the moment, such that time is measured in units of $\hbar\Gamma^{-1}$. The experimental literature on semiconductor QDs indicate~\cite{simmons2009charge} that it is reasonable to assume $U\sim 20\Gamma$ and $V\sim 10\Gamma$. Moreover, recent experimental developments have demonstrated that it is possible to tune the parameters of arrays of quantum dots to specific purposes~\cite{yu2016tunable, hensgens2017quantum}. Tipically, the charging energy $U$ can be controlled by the size of the quantum dot, and the capacitive coupling $V$ can be controlled by the distance between neighboring quantum dots. The tunneling coupling can be adjusted by a combination of gate voltages and the distance between quantum dots. The single-electron energies $\varepsilon_l$ can be adjusted via gate voltages. Starting from the previous set of parameters we fixed $V\sim 10\Gamma$, $\epsilon_l=0$ for $l=0,1,2$, and found by varying $U$ that we could get much better gate fidelity by using $U=21.83\Gamma$, as shown in Fig. \ref{fig:fidelity}. We use this set of parameters for the remainder of our work.

With this model there are nine basis states in the $N=2$ subspace and another nine in the $N=4$ subspace. In both cases, only three of them are logical states: $\ket{001}$, $\ket{010}$ and $\ket{100}$ for $N=2$, and $\ket{110}$, $\ket{101}$ and $\ket{011}$ for $N=4$. All remaining basis states correspond to leakage states where some quantum dots become populated by a single electron. In Fig. \ref{fig:swap} we plot the time-evolution of each of these six logical states under Hamiltonian \ref{eq:hamiltonian} using $U=21.83\Gamma$, $V=10\Gamma$ and $\epsilon_l=0$ for $l=0,1,2$.

We now analyze the dynamics in the subspace $\mathcal{H}_1$, where the control is set to $c=1$. We assume the single-particle energies are identical for the two target QDs, $\varepsilon_1=\varepsilon_2=0$. If we prepare the joint state of the target QDs in
\begin{equation}
\ket{\psi_{c=1}(0)} = c^\dagger_{1\uparrow}c^\dagger_{1\downarrow}\ket{\emptyset}_1\ket{\emptyset}_2 = \ket{1}_1\ket{0}_2,
\end{equation}
it evolves to the superposition 
\begin{equation}
\begin{alignedat}{2}
\ket{\psi_{c=1}(t)} 
=\quad&f_{10}(t)\ket{1}_1\ket{0}_2\\
+~&f_{01}(t)\ket{0}_1\ket{1}_2\\
+~&f_L(t)(c^\dagger_{1\uparrow}c^\dagger_{2\downarrow}+e^{i\theta}c^\dagger_{2\uparrow}c^\dagger_{1\downarrow})\ket{\emptyset}_1\ket{\emptyset}_2.
\end{alignedat}
\end{equation}
All coefficients are oscillatory functions of time, as seen in Fig.~\ref{fig:swap}: $f_{01}$ and $f_{10}$ have two superimposed oscillatory components with distinct frequencies, 
\begin{align}
&\Omega_1=\sqrt{16+(U-V)^2}, 
\\
&\Omega_2=\frac{U-V+\Omega_1}{2},
\\
&\Omega_3=\frac{U-V-\Omega_1}{2}.
\end{align}
Ideal gate completion is achieved when $|f_{01}|^2=1$. This can only happen at the maximum of the slowest varying component of $f_{01}(t)$, that is, when $t^*=2\pi/\Omega_3$, leading to the Fredkin gate time shown in the main text in SI units:
\begin{equation}
t^*= \frac{2\pi\hbar}{|U-V-\sqrt{16\Gamma^2+(U-V)^2}|}.
\end{equation}
In general, the higher frequency components $\Omega_1$ and $\Omega_2$ will not have maxima exactly at the same time, precluding ideal gate performance. However, for $U\gg V$ (as it is the case in actual QDs devices) there are always high frequency maxima close enough to $t^*$ to allow for 
$\left| |f_{01}(t^*)|^2-1\right| \lesssim 10^{-3}$.

As we mentioned previously, the dynamics imposed by $H_F$ leads to the appearance of components in $\ket{\psi(t)}$ which do not belong to the computational basis. This could, in principle, lead to leakage errors, measured by $|f_L(t)|^2$. It can be shown, however, that, if the target QDs are identical,
\begin{equation}
|f_L(t)|^2 = \frac{2(1-\cos\Omega_1 t)}{\Omega_1^2}.
\end{equation}
Once again, for $U\gg V$ the amplitude of $|f_L(t)|^2$ is small and its minima are closely spaced enough so that $|f_L(t^*)|^2\approx 0$, as seen in Fig.~\ref{fig:swap} for $U=21.83$ and $V=10$.

If we now set  $a=1$, $b=0$, and set the control bit to 0 closed form analytical solutions for the dynamics are too lengthy to be useful. Nevertheless, numerical solutions, such as the
one depicted in Fig~\ref{fig:swap}, show that the probability for the system to remain in the initial state oscillates slightly below 1. For parameter values representative of real
semiconductor QDs, this probability is larger than 0.99 at all times.

\subsection{Quasistatic noise}

We now consider how the performance of the gate is affected by electrostatic fluctuations that change each quantum dot's single-particle energies. First we consider low-frequency (``quasistatic'') charge noise resulting from electrostatic fluctuations that are slow enough to be considered approximately constant during a single gate operation. We incorporate these in our model through variations of $\varepsilon_l$ from one gate operation to the other. Studying the eigenstates of the Hamiltonian, we found numerically that the probability for gate achievement in the case where the control bit is set to 1 depends on $\varepsilon_l$ as 
\begin{equation}
p(\varepsilon_1,\varepsilon_2) =\left[ f_0-\Lambda\left(\frac{\varepsilon_1-\varepsilon_2}{\Gamma}\right)^2\right]^2,
\end{equation}
where 
\begin{equation}
\Lambda\equiv \alpha\left(\frac{U-V}{\Gamma}\right)^2 
+ \beta,
\end{equation}
$f_0$ is the fidelity in the absence of noise, $|f_0-1|\lesssim 10^{-3}$, $\alpha\approx 0.12$, and $\beta \approx 0.25$. Assuming that $\varepsilon_l$ are independent gaussian random variables with zero
average and $\langle\varepsilon^2_l\rangle = \bar{\varepsilon}^2$, the average of $p$ is given by
\begin{equation}
\langle p\rangle = \left[f_0 - 2\Lambda\left(\frac{\bar{\varepsilon}}{\Gamma}\right)^2\right]^2 +
 8\Lambda^2\left(\frac{\bar{\varepsilon}}{\Gamma}\right)^4.
\end{equation}
To obtain a rough estimate of the effects of quasistatic noise, using the previous values of $U=21.83\Gamma$ and $V=10\Gamma$, and assuming $\bar{\varepsilon}\sim 0.01\Gamma$, we get
\begin{equation}
\langle p\rangle - p_0 \sim 4.6\times 10^{-2}
\end{equation}
where $p_0$ is the probability of gate achievement in the absence of noise.

\subsection{High frequency noise}

\begin{figure}
\centering
\includegraphics[width=0.9\columnwidth]{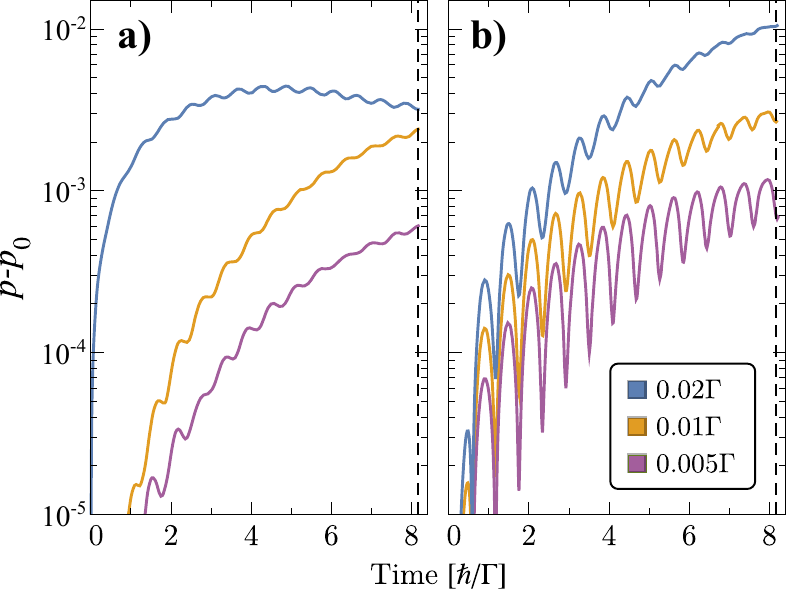}
\caption{\textbf{High-frequency noise.} \textbf{a)} Change in the success probability of the Fredkin gate due to high-frequency noise and respective standard deviation (\textbf{b)}). We considered noise amplitudes of 0.005$\Gamma$, 0.01$\Gamma$, and 0.02$\Gamma$, averaged over 1000 runs of the gate.}
\label{fig:5}
\end{figure}

We also consider the possibility of electrostatic fluctuations that change the single-particle energies during the execution
of the gate. We model those fluctuations as gaussian noise superimposed to the QD's potentials,
\begin{equation}
W = \sum_{l\sigma}X_l(t)a^\dagger_{l\sigma}a_{l\sigma}
\end{equation}
where $X_l(t)$ are independent gaussian random variables, which we assumed to have identical distributions. The time evolution is now given by a system of stochastic differential 
equations that need to be solved numerically. We employed the same values of parameters previously used for assessing the effect of quasistatic noise: $U=21.83\Gamma$, $V=10\Gamma$, 
$X=0.01\Gamma N(0,1)$, where $N(0,1)$ is a gaussian random variable with zero average and unit variance. After averaging over 1000 runs of the gate we find that the change in success probability at the time for gate completion is slightly larger than $10^{-3}$ and its standard deviation in the same order, shown in Figure \ref{fig:5}.

\subsection{Single-electron encoding}

We will now address the consequences of adopting the single-electron encoding, corresponding to associating the presence of a single electron in a QD with the logical state $\ket{0}$ and an empty QD with the logical state $\ket{1}$. When compared with the double occupancy encoding, the single-electron version may be more amenable to experimental implementation, since here the swap operation comprises the hopping of a single electron from one QD to another. In contrast, the swap operation in the double occupancy encoding corresponds
to the coherent tunneling of a pair of electrons from one QD to another. This process is slower and more susceptible to the effects of dephasing, induced by energy relaxation. 

To avoid leakage into states outside the computational basis, in the single-electron encoding it is necessary to fix the spin of the electrons. By applying a sufficiently strong static and uniform magnetic field one can guarantee that the electrons loaded into the QDs will have their spins locked anti-parallel to the applied field. In this case, the Pauli exclusion
principle precludes double occupancy of the target QDs, thus avoiding leakage. Thus, from the four possible initial states of the target bits, only two will have non-trivial dynamics: 
\begin{equation}
|1\rangle_1|0\rangle_2\equiv c^\dagger_{2\uparrow}|\emptyset\rangle_1|\emptyset\rangle_2,  
\ \  
|0\rangle_1|1\rangle_2\equiv c^\dagger_{1\uparrow}|\emptyset\rangle_1|\emptyset\rangle_2.
\end{equation}
As in the double-occupancy encoding, the presence of an electron in the control QD will have the effect of an imbalance between the on-site energies of the two target QDs, through the capacitive interaction between the control QD and the first target QD. This will suppress the hopping between QD1 and QD2 with high probability. If the control QD is empty, however, the on-site energies of the two target QDs will be aligned and resonant hopping will take place with probability 1 within a time $\pi\hbar/2\Gamma$. Thus, gate completion in the single-electron implementation can be almost six time faster than in the double-occupancy encoding.

\subsection{Role of energy relaxation}

Our proposal for the implementation of a logical reversible gate using quantum dots rely on the coherent nature of the underlying quantum dynamics. We have shown in previous sections that the dynamics we proposed is robust against two common sources of noise in quantum dots, low-frequency and high-frequency electrostatic noise. Another issue relevant for the viabiality of our proposal is energy relaxation, as produced, for instance, by electron-phonon coupling. It is crucial that logical operations that rely on coherent quantum dynamics are performed faster than typical energy relaxation times. Our estimates for gate speeds, based on typical semiconductor quantum dot parameters, are in the range of hundreds of picoseconds for the double-occupancy enconding and tens of picoseconds for the single-electron encoding. Recent experimental studies of coherent quantum dynamics have reported energy relaxation time in the nanoseconds, indicating that the gate implementation we propose here is fast enough to beat decoherence induced by energy relaxation.~\cite{PRApp9-18}

\subsection{Experimental feasibility}

\begin{figure}
\centering
\includegraphics[width=0.8\columnwidth]{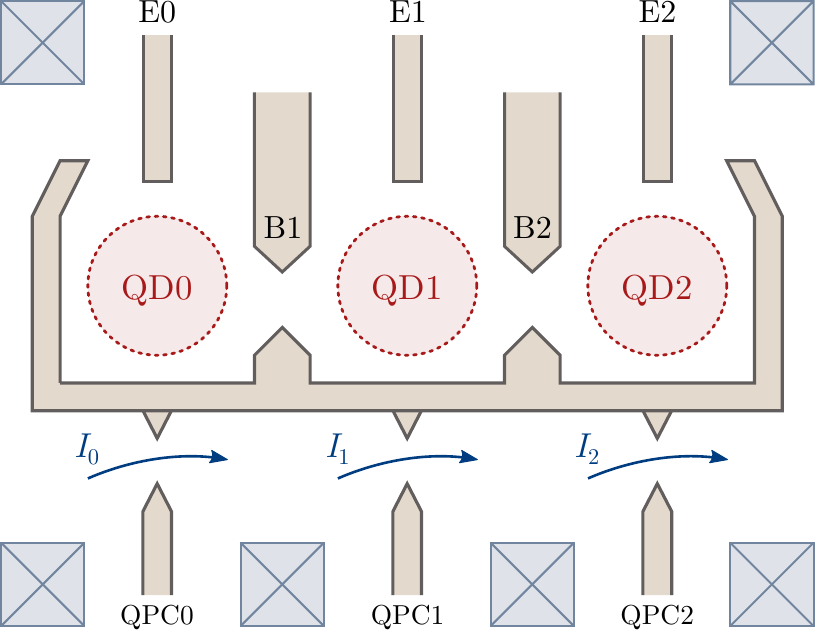}
\caption{\textbf{Electrostatic gate organization for a linear triple-dot array.} Scheme for an implementation of our Fredkin gate design with semiconductor quantum dots on a GaAs/AlGaAs substrate. The quantum dots are defined by voltages applied to TiSi gates. Voltages on contacts E0, E1 and E2 control the individual qubits' on-site potentials; voltages on contacts B1 and B2 regulate the tunneling and capacitive interaction between the quantum dots; the three quantum point contacts QPC0, QPC1 and QPC2 are used to measure the charge state in each quantum dot through the currents $I_0, I_1, I_2$. Further simplifications may be considered, such as using a single QPC which may be able to measure the charge state of all three dots.}
\label{fig:2}
\end{figure}

We now describe an example of a concrete implementation of our Fredkin gate design with three quantum dots in semiconductors. The design takes into account the properties and limitations of current experimental techniques, and is believed to be feasible with nowadays technology. An example of such experimental techniques can be found in \cite{Schroer2007}.

The quantum dots are realized on a semiconductor substrate composed by a layer of GaAs a few micrometers thick, upon which a layer of doped GaAs (or AlGaAs) is deposited. The matching of the chemical potential at the interface between the two different layer induces a two-dimensional electron gas (2DEG), of density $\simeq 10^{11}\text{e}/\text{cm}^2$ and mobility $\simeq 75 \text{m}^2/\text{V s}$. On top of the sample, $\simeq 100$ nm above the 2DEG layer, the structure of the Ti/Au gates is defined through electron beam lithography. With reference to Figure \ref{fig:2}, the defined structure comprises the following elements:
\begin{itemize}
	\item Main confinement structure: an applied voltage defines the potential confinement of the quantum dots on the underlying 2DEG layer through the generated Coulomb potential. 
	\item Ohmic contacts, represented as the square crossed boxes.
	\item Individual quantum dot control electrodes, E0, E1 and E2: applied voltages allow the control of each individual dot's on-site potential $\varepsilon_0$, $\varepsilon_1$ and $\varepsilon_2$.
	\item Tunnelling barrier control electrods B1 and B2: voltages applied here allow for the control of the tunneling couplings $\Gamma$ and $\Gamma^*$. By suitable tuning of the potentials, the device can be configured from three completely isolated quantum dots, all the way to a single confinement region with no separation between the three parts.
	\item Quantum point contacts QP0, QP1, QP2: functioning as charge point contacts, they allow for the measurement of presence of charge in each quantum dot, thus measuring the dot's logical state.   
\end{itemize}  
In a normal operational regime, the device requires electrons to flow through it from QD0 through QD1 to QD2, from which they can then leave the device. This is the main flow, required for the initialization of each dot's state prior to operation, and for the discharging of the dots after the operation. 

During the initialization stage, suitable operation of the main channel bias voltage Vb and of the voltages applied to B1, B2, E0, E1 and E2 allows for the charging of each dot with the required number of electrons. 

After the initialization is completed, the voltages are tuned as to implement the desired interaction Hamiltonian as described previously in Section \ref{sec:fredkingate}.

Additional electronic flows are required for the operation of the quantum point contacts QP0, QP1 and QP2 during the measurement of the charge present in each dot.

The capacity to control the quantum dots at the level of single electrons relies heavily on the ability to keep thermal noise low and to increase the capacitance of each dot, in order to increase the Coulomb blockade effect. Such requirements can be met if the sample is maintained at a temperature of the order of 300mK, achievable with He$^4$-He$^3$ dilution refrigerators, and due to the capacitance dependence on the temperature, the size of the quantum dots can be taken up to as large as 100nm. 


\begin{table}[t]
\centering
\begin{tabular}{ccc|cc|c}
\hline\hline
~$p$~ & ~$q$~ & ~$r$~ & \textit{carry} & \textit{parity} &  ~Fidelity~\\\hline
~0~ & ~0~ & ~0~ & ~0~ & ~0~ & ~0.986~\\
~0~ & ~0~ & ~1~ & ~0~ & ~1~ & ~0.991~\\
~0~ & ~1~ & ~0~ & ~0~ & ~1~ & ~0.994~\\
~0~ & ~1~ & ~1~ & ~1~ & ~0~ & ~0.997~\\ 
~1~ & ~0~ & ~0~ & ~0~ & ~1~ & ~0.994 ~\\
~1~ & ~0~ & ~1~ & ~1~ & ~0~ & ~0.997 ~\\
~1~ & ~1~ & ~0~ & ~1~ & ~0~ & ~0.994 ~\\
~1~ & ~1~ & ~1~ & ~1~ & ~1~ & ~0.999 ~\\
\hline\hline
\end{tabular}
\caption{Full-adder truth table with the fidelities obtained for each possible initial state of the full-adder, for parameter values $U=21.83\Gamma$ and $V=10\Gamma$.}
\label{table:fidelity1}
\end{table}

\section{Full-Adder Fidelity}

In this section we present the fidelities obtained by simulating the full-adder protocol discussed in the main text. Only three parameters need to be controlled during the protocol: $\Gamma$, $\Gamma^*$ and $\varepsilon_0$. The simulations were done with the following considerations:
\begin{itemize}
\item We simulated each step independently, thus considering that at the beginning of each step there is an ideal measurement on QD0 that collapses the wave function to the correct logic state.
\item We considered that the fidelity for each step is the value $|c|^2$ corresponding to the amplitude of the correct logic state for that step.
\item We considered that the fidelity for one run of the full adder is the product of the fidelities of all intermediate steps.
\end{itemize}

In Table \ref{table:fidelity1} we show the fidelities obtained for the full adder using the parameters we have been considering so far: $U=21.83\Gamma$ and $V=10\Gamma$. The fidelity obtained is above $0.99$ for 7 of the 8 initial states, and slightly below for the $(p,q,r) = (0, 0, 0)$.

\section{Measurement Cost}

Measuring the charge state of quantum dots can be done through different techniques. The most common is through the usage of a Quantum Point Contact (QPC), which is created by an electrostatic gate neighbouring a quantum dot, as illustrated in Fig. \ref{fig:2}. The conductance in the QPC is highly sensitive to changes in the surrounding electric field. As such, electrons flowing in or out of the quantum dot lead to abrupt changes in the current flowing through the QPC. As an example application of a QPC, in \cite{vandersypen2004real}, the authors demonstrate experimentally the measurement of single-electron fluctuations in a quantum dot. In their setup they are able to detect electron changes in the quantum dot in less than 10 $\upmu$s by biasing the QPC with $1$ mV voltage and using a current of $30$ nA. In order to obtain a simple estimate of the energetic cost of this operation we can write
\begin{equation}
\Delta E_M = VI\Delta t,
\end{equation}
which leads to a value of $\Delta E_M = 3.6$ keV using the mentioned values.

\end{document}